\documentclass{PoS}
\newcommand{\Slash}[1]{{\ooalign{\hfil/\hfil\crcr$#1$}}}

\title{The OPE of the B-meson light-cone wavefunction for exclusive B decays: radiative corrections and higher-dimensional operators}

\ShortTitle{The OPE of the B-meson light-cone wavefunction for exclusive B decays
}

\author{Hiroyuki Kawamura\\
        Radiation Laboratory, RIKEN, Wako 351-0198, Japan\\
        E-mail: \email{hiroyuki@ribf.riken.jp}}

\author{Jiro Kodaira%
\thanks{Deceased.}\\
        Theory Division, KEK, Tsukuba 305-0801, Japan
}

\author{\speaker{Kazuhiro Tanaka}%
\\
       Department of Physics, Juntendo University,
    Inba, Chiba 270-1695, Japan\\
        E-mail: \email{tanakak@sakura.juntendo.ac.jp}}

\abstract{We discuss the $B$-meson light-cone wavefunction relevant for
QCD factorization approach for exclusive $B$-meson decays.
We derive the operator product expansion for the $B$-meson light-cone wavefunction, 
taking into account the local composite operators of dimension less than 6
and calculating the radiative corrections at order $\alpha_s$ 
for the corresponding Wilson coefficients.
The result embodies peculiar UV and IR behaviors 
of the $B$-meson light-cone wavefunction,
the Sudakov-type double logarithmic effects and 
the mixing of the multiparticle states 
with additional 
gluons inside the $B$ meson.
The former effects are
induced from the cusp singularity in the
radiative corrections, while the latter
is manifested by the participation of the higher-dimensional operators
associated with the nonperturbative structure of the $B$ meson. 
}

\FullConference{8th International Symposium on Radiative Corrections \\
		 October 1-5, 2007\\
		 Florence, Italy}

\begin{document}
For the exclusive $B$-meson decays such as $B \to \pi \pi$, 
$\rho \gamma, \ldots$, systematic methods have been developed
using QCD factorization based on the heavy-quark limit~\cite{Beneke:2000ry}.
Among the building blocks in the corresponding factorization formula of the decay amplitude,
essential roles are played by
the light-cone 
wavefunctions (LCWFs) for the participating mesons, 
which include
nonperturbative long-distance contribution.
In particular, not only the LCWFs for the light mesons $\pi, \rho, \ldots$,
produced in the final state,
but also those for the $B$ meson participate in 
%
processes where large momentum is transferred to the soft 
spectator quark via hard gluon
exchange~\cite{Beneke:2000ry}.
We define the leading quark-antiquark component of 
$B$-meson LCWF 
as the vacuum-to-meson matrix element~\cite{Grozin:1997pq}: 
\begin{equation}
\tilde{\phi}_B(t, \mu)
=
\langle 0|
\bar{q}(tn)
{\rm P}e^{ig\int_0^td\lambda n\cdot A(\lambda n)}
n \hspace{-0.45em}/
\gamma_5h_v(0) 
|\bar{B}(v)\rangle 
=\int d\omega e^{-i\omega t}
\phi_B(\omega, \mu) .
\label{eq1}
\end{equation}
Here the bilocal operator 
is built of 
the $b$-quark and light antiquark 
fields, $h_v(0)$ and $\bar{q}(tn)$, linked by the Wilson line
at a light-like 
separation 
$tn$ with $n_\mu$ the light-like vector 
($n^2 =0$, $n\cdot v=1$), $v_\mu$ is
4-velocity of the $B$ meson, and the RHS defines 
the momentum representation 
with $\omega v^+$ denoting the LC component of the light antiquark's momentum. 
A difference compared with the familiar pion-LCWF is that $h_v(0)$ 
is an effective field in the heavy-quark effective theory (HQET)~\cite{Neubert:1994mb}.
In (\ref{eq1}), $\mu$ denotes
the scale where the bilocal operator is renormalized.
If 
one could Taylor expand the bilocal operator,
the WF (\ref{eq1}) would be equivalent to matrix element 
of a set of local operators of the type,
$\bar{q}(0)(n\cdot D )^j
n \hspace{-0.45em}/
\gamma_5 h_v (0)$.
Indeed, for the light mesons $\pi, \rho, \ldots$,
such correspondence holds and is useful for establishing
systematic framework to calculate
the LCWFs explicitly~\cite{Braun:1990iv}. 
In contrast, 
for the $B$-meson LCWF (\ref{eq1}),
such straightforward correspondence is broken as demonstrated below,
because the twist is not a good quantum number by presence
of the effective heavy-quark field.

The ``IR structure'' of the $B$-meson LCWF has been studied~\cite{KKQT}
using constraints from equations of motion (EOM), 
$\bar{q}\overleftarrow{\Slash{D}}  = v \cdot \overrightarrow{D} h_v  = 0$,
and heavy-quark symmetry, $\Slash{v}h_v=h_v$.
These constraints are solved to give (\ref{eq1})
as the sum of the 
two terms,
$\phi_B (\omega ) = \phi_B^{(WW)} (\omega ) + \phi_B^{(g)} (\omega )$.
The first term is expressed by the analytic formula as 
$\phi_B^{(WW)} (\omega ) = iF\omega \theta (2\bar{\Lambda}  - \omega )/( 2\bar{\Lambda}^2 )$,
in terms of the HQET parameters~\cite{Neubert:1994mb}, 
the decay constant $F=-i\langle 0|
\bar{q}
n \hspace{-0.45em}/
\gamma_5h_v 
|\bar{B}(v)\rangle$
and the ``effective mass'' $\bar{\Lambda}=m_B -m_b$ representing the mass difference
between $B$-meson and $b$-quark. The second term $\phi _B^{(g)}$ is obtained as a certain 
integral
of matrix element of 
3-body LC operator,
$\langle 0 |\bar{q}(tn)G_{\alpha \beta}(un)\sigma_{\lambda \eta} h_v (0)| {\bar B(v)} \rangle$,
where\footnote{
The Wilson lines connecting the constituent fields are suppressed for simplicity.}
the nonperturbative 
gluons generated inside the $B$ meson participate as the field strength tensor $G_{\alpha \beta}$.
On the other hand, the ``UV structure'' of the $B$-meson 
WF has been studied in \cite{Lange:2003ff}
by calculating the 1-loop renormalization of the bilocal operator in (\ref{eq1}).
The ``vertex-type'' correction around a ``cusp'' between the two Wilson lines,
the light-like one of (\ref{eq1}) and the time-like one from 
$h_v(0)={\rm P}\exp [ig\int_{-\infty}^0d\lambda v\cdot A(\lambda v)]h_v(-\infty v)$,
produces the ``radiation tail'' 
as $\phi_B(\omega, \mu) \sim -iFC_F \alpha_s \ln(\omega/\mu)/(\pi \omega)$
for $\omega \gg \mu$, where $C_F = (N_c^2-1)/(2N_c)$.
This implies that the moments $\int_0^\infty  d\omega  \omega ^j \phi _B (\omega, \mu)$,
which would correspond to matrix element of the local operators mentioned above,
are divergent
reflecting the singularity associated with the cusp
of the two Wilson lines~\cite{Lange:2003ff}.

To see the behavior of the $B$-meson WF incorporating both IR and UV 
structures, 
we calculate the radiative corrections 
taking into account hard and soft/collinear loops.
We calculate the 1-loop diagrams for the 2-point function (\ref{eq1}) 
in $4-2\varepsilon$ dimensions.
We first note that the similar diagrams for the case of the pion LCWF,
defined by
the light-$q \bar{q}$ bilocal operator,
yield
the ``cancelling'' UV and IR poles, $1/\varepsilon_{UV}-1/\varepsilon_{IR}$,
reflecting the scaleless loop-integral in massless QCD,
and this structure is accompanied 
by the convolution with 
the splitting function
(Brodsky-Lepage kernel)~\cite{Braun:1990iv};
the corrections are analytic (Taylor expandable) at $t=0$.
Now, for the $B$-meson LCWF (\ref{eq1}), we get\footnotemark[1]
\begin{eqnarray}
\lefteqn{\!\!\!\!\!\!\!\!\!\!\!\!\!\!\!
\tilde \phi_B^{\scriptsize \mbox{1-loop}} (t,\mu ) 
=
\frac{\alpha_s C_F }{2\pi }
\int_0^1 {d\xi } \left[ \left\{  - \left( \frac{1}{2\varepsilon_{UV}^2} 
+ \frac{L}{\varepsilon_{UV}} + L^2+ \frac{5\pi^2 }{24} \right)
\delta (1 - \xi ) 
 + \left( \frac{1}{\varepsilon_{UV}} - \frac{1}{\varepsilon_{IR}} \right)
\left[ \frac{\xi}{1 - \xi} \right]_+
\right. \right.}
\nonumber\\   
&&\left.\left.   \!\!\!\!\!\!
-
\left( \frac{1}{2\varepsilon_{IR} } 
+ L \right) \right\}
\langle 
\bar{q}(\xi tn) n \hspace{-0.45em}/ \gamma_5 h_v (0) 
\rangle  
  - t\left( \frac{1}{\varepsilon_{IR}} + 2 L- 1 - \xi  \right) 
\langle 
\bar{q}(\xi tn)v \cdot \overleftarrow{D}  n \hspace{-0.45em}/ \gamma_5 h_v (0)
\rangle  
\right] + \cdots ,
\label{eq2}
\end{eqnarray}
where $L\equiv \ln(it \mu e^{\gamma_E})$
with the ${\overline{\rm MS}}$ scale $\mu$,
and $\langle \cdots \rangle \equiv \langle 0| \cdots|\bar{B}(v) \rangle$.
The ``vertex-type'' correction 
that connects the light-like Wilson line and the $\bar{q}(tn)$ field
in (\ref{eq1})
is associated with only the massless degrees of freedom, and 
gives the term with the ``plus''-distribution $[\xi/(1-\xi) ]_+$, which
has the same structure as the corresponding correction in the pion case.
But the other diagrams give novel behavior
with ``non-cancelling'' UV and IR poles:
the another vertex-type correction gives~\cite{Lange:2003ff} the
terms proportional to 
$\delta (1-\xi)$, which contain
the double as well as single UV pole, corresponding to the cusp singularity mentioned above.
The ``ladder-type'' correction, connecting the two quark fields in (\ref{eq1}), gives
the last line in (\ref{eq2}), which contains
the IR poles and is associated with not only the bilocal operator in (\ref{eq1}) 
but also the higher dimensional operators;
note that the ellipses in (\ref{eq2}) are expressed by those operators
involving the 
two or more additional
covariant derivatives. 

The renormalized LCWF, obtained by subtracting the UV poles from (\ref{eq2}),
is non-analytic at $t=0$ by presence of logarithms $L$, $L^2$~\cite{Lange:2003ff,Braun:2003wx}.
In particular, the nontrivial dependence on $t\mu$ through $L$
implies that the scale $\sim 1/t$ plays a role to separate the UV and IR regions.
Thus we have to use the operator product expansion (OPE) to
treat the different UV and IR behaviors simultaneously:
the coefficient functions absorb all the singular logarithms,
while, for the local operators to absorb the IR poles, we have to take into account many 
higher dimensional operators.
Such OPE with local operators is useful when the separation $t$
is less than the typical distance scale of quantum fluctuation,
i.e., when $t\lesssim 1/\mu$.
The OPE result can be 
evolved to a higher scale
by the evolution operator associated with the single-UV-pole terms in (\ref{eq2}),
i.e., by the Brodsky-Lepage-type kernel and the Sudakov-type operator 
with the cusp anomalous dimension~\cite{Lange:2003ff,Braun:2003wx}.

An OPE for the $B$-meson LCWF (\ref{eq1}) 
has been discussed 
in \cite{Lee:2005gza},
taking into account the local operators of dimension less than 5
and 
the NLO ($O(\alpha_s)$)
corrections to the corresponding Wilson coefficients 
in a 
``cutoff scheme'', where
specifically an additional momentum cutoff $\Lambda_{UV}$ is introduced 
and the OPE is derived for the regularized moments,
$\int_0^{\Lambda_{UV}} d\omega  \omega^j \phi_B (\omega, \mu)$ with $j=0,1$.
In this work we derive the OPE for the $B$-meson LCWF (\ref{eq1}),
taking into account the local operators of dimension less than 6
and calculating the radiative corrections for the corresponding Wilson coefficients 
at NLO accuracy.
Following the above discussion, we carry out the calculation 
in the $\overline{\rm MS}$ scheme for $t\lesssim 1/\mu$, 
so that there is no need to introduce any additional cutoff.
%

The most complicated part of this calculation 
is to reorganize the contributions from a gauge-invariant 
set of Feynman diagrams 
in terms of 
matrix element
of 
gauge-invariant operators including higher dimensional ones.
In particular, to derive the Wilson coefficients associated with the dimension-5
operators such as $\bar{q}G_{\alpha\beta}n \hspace{-0.45em}/ \gamma_5 h_v$, 
we have to compute the 1-loop diagrams for the 3-point function,
as well as those for the 2-point function as in (\ref{eq2}),
where the former diagrams are obtained by attaching the external gluon line
to the latter diagrams in all possible ways. 
To perform the calculation in a systematic and economic way, we employ the background 
field method~\cite{Abbott:1980hw}.
We decompose the quark and gluon fields into the ``quantum'' and ``classical'' parts,
where the latter part represents the nonperturbative long-distance 
degrees of freedom inside $B$ meson as a background field and satisfies the classical EOM.
The quark and gluon propagators for the quantum part contain 
the coupling with an arbitrary number of background fields, and 
each building block of the Feynman diagrams obeys the exact 
transformation property under the gauge transformation for the background fields.
We use the Fock-Schwinger gauge, $x^\mu  A_\mu^{(c)} (x)=0$,
for the background gluon field $A_\mu^{(c)}$.
This gauge condition is solved to give 
$A_\mu^{(c)}(x) = \int_0^1 {du} u x^\beta  G_{\beta \mu}^{(c)} (ux)$~\cite{Abbott:1980hw},
which allows us to reexpress each Feynman diagram in terms of
matrix element of the operators associated with the field strength tensor.

Using this framework, 
the tree-level matching to derive our OPE
can be performed replacing each constituent
field in 
(\ref{eq1}) 
by the corresponding background
fields.
The {\it classical} bilocal operator can be Taylor expanded, 
and
we obtain the OPE at the tree level
with the local operators of dimension-3, -4, and -5 (see (\ref{eq3}) below).

For the 1-loop matching, we calculate the 1-loop corrections to 
the 2- and 3-point functions with the insertion of the bilocal operator in (\ref{eq1}),
taking into account the mixing of operators of dimension less than 6.
Apparently the mixing through the UV region of the loop momenta 
arises only in the 2-point function,
whose result can be immediately obtained from (\ref{eq2}); but the additional
mixing can arise
in both 2- and 3-point functions accompanying the IR poles.
We perform the loop integration in the coordinate space 
using the Schwinger (``heat kernel'') representation 
of the Feynman amplitudes under the background fields~\cite{Abbott:1980hw}.
For the calculation of the 3-point function,
the Fock-Schwinger 
gauge ensures that the Wilson line in (\ref{eq1}), 
as well as the heavy-quark field,
does not couple directly to the background gluons.
We reorganize the result in terms of a complete set of gauge-invariant operators
using the EOM and heavy-quark symmetry.
We subtract the UV poles to renormalize the bilocal operator of (\ref{eq1}),
and also renormalize the local operators to absorb the IR poles.
Combining this result with the above tree-level result,
we obtain the OPE for the bilocal 
operator in (\ref{eq1}),  
$\bar{q}(tn)
{\rm P}e^{ig\int_0^td\lambda n\cdot A(\lambda n)}
n\hspace{-0.45em}/\gamma_5h_v(0)
=\sum_{i}C_i(t,\mu) {\cal O}_i(\mu)$,
to the desired accuracy as
\begin{eqnarray}
\lefteqn{ \bar{q}(tn)
{\rm P}e^{ig\int_0^td\lambda n\cdot A(\lambda n)}
n\hspace{-0.45em}/\gamma_5h_v(0)
=
\left[1-  \frac{\alpha_sC_F}{4\pi}
\left(2L^2+2L+ \frac{5\pi^2}{12}\right)\right]{\cal O}^{(3)}_1}
\nonumber\\&&
\;\;\;\;\;\;\;\;\;\;\;\;\;\;\;\;\;\;\;\;\;\;\;\;\;\;\;\;\;\;\;\;\;\;
-it\left\{
\left[
1- \frac{\alpha_sC_F}{4\pi}
\left(2L^2+L+\frac{5\pi^2}{12}\right)
\right] {\cal O}^{(4)}_1
-\frac{\alpha_sC_F}{4\pi}\left(4L-3\right){\cal O}^{(4)}_2
\right\}
\nonumber\\&&
\;\;\;\;\;\;\;\;\;\;\;\;\;\;\;\;\;\;\;
-\frac{t^2}{2}\left\{
\left[1- \frac{\alpha_sC_F}{4\pi}
\left(
2L^2+\frac{2}{3}L+\frac{5 \pi^2}{12}
\right)\right]{\cal O}^{(5)}_1
\right.
-\frac{\alpha_sC_F}{4\pi}
\left(4L-\frac{10}{3}\right)\left( {\cal O}^{(5)}_2
+ {\cal O}^{(5)}_3 \right)
\nonumber\\&&\;\;\;\;\;\;\;\;\;\;\;\;\;\;\;
+ 
\frac{\alpha_s}{4\pi}
\left[ C_F\left(-4L+\frac{10}{3}\right)
+ C_G\left(7L-\frac{13}{2}\right)
\right]{\cal O}^{(5)}_4
+ 
\frac{\alpha_s}{4\pi} \left(-\frac{4}{3}C_F +C_G \right) \left(L-1\right)
{\cal O}^{(5)}_5
\nonumber\\&&\;\;\;\;\;\;\;\;\;\;\;\;\;\;\;
+
\frac{\alpha_s}{4\pi} \left(-\frac{2}{3}C_F +C_G \right) \left(L-1\right)
{\cal O}^{(5)}_6
\left.
+ 
\frac{\alpha_s}{4\pi}\left(-\frac{1}{3}C_F +\frac{1}{4}C_G \right) \left(L-1\right)
{\cal O}^{(5)}_7
\right\},
\label{eq3}
\end{eqnarray}
where $C_G=N_c$ and a basis of local operators, ${\cal O}^{(d)}_k$ 
($k=1,2,\ldots$), of dimension-$d$ is defined as
${\cal O}^{(3)}_1\equiv \bar{q}n\hspace{-0.45em}/\gamma_5h_v$,
$\{ {\cal O}^{(4)}_k \}\equiv\{ \bar{q}(i n\cdot \overleftarrow{D})
n\hspace{-0.45em}/\gamma_5h_v$,
$\bar{q}(iv\cdot \overleftarrow{D})
n\hspace{-0.45em}/\gamma_5h_v\}$, 
and
$\{ {\cal O}^{(5)}_k \}\equiv\{ 
\bar{q}(in\cdot \overleftarrow{D})^2
n\hspace{-0.45em}/ \gamma_5 h_v$,
$\bar{q}(iv\cdot \overleftarrow{D})
(in\cdot \overleftarrow{D})n\hspace{-0.45em}
/\gamma_5 h_v$,
$\bar{q}(iv\cdot \overleftarrow{D})^2
n\hspace{-0.45em}/\gamma_5 h_v$,
$\bar{q}igG_{\alpha\beta}v^\alpha n^\beta n\hspace{-0.45em}/
\gamma_5h_v$,
$\bar{q}igG_{\alpha\beta}\gamma^{\alpha}n^{\beta}
\bar{n}\hspace{-0.45em}/\gamma_5h_v$,
$\bar{q}igG_{\alpha\beta}\gamma^\alpha v^\beta
\bar{n}\hspace{-0.45em}/
\gamma_5h_v$,
$\bar{q}gG_{\alpha\beta}\sigma^{\alpha\beta}
n\hspace{-0.45em}/\gamma_5h_v\}$.
Here we have introduced another light-like vector, $\bar{n}^2=0$,
as $v_\mu=(n_\mu+\bar{n}_\mu)/2$. 
The double logarithm $L^2$ in the coefficient functions
originates from cusp singularity.
The 1-loop corrections for the 3-point function 
induce only the operators ${\cal O}^{(5)}_{4,5,6,7}$
associated with the field strength tensor, while those for the 2-point function
induce all ten operators through the use of the EOM.

Taking the matrix element $\langle \cdots \rangle \equiv \langle 0| \cdots|\bar{B}(v) \rangle$
of (\ref{eq3}), we can derive the OPE form of 
the $B$-meson LCWF (\ref{eq1}).
Matrix elements of the local operators appearing 
in the RHS of (\ref{eq3}) are known to be related to a few nonperturbative parameters
in the HQET, using the EOM and heavy-quark symmetry as demonstrated in
\cite{Grozin:1997pq,KKQT}:
$\langle {\cal O}^{(4)}_1 
\rangle= 4iF  \bar{\Lambda}/3$, 
$\langle {\cal O}^{(4)}_2 
\rangle
= iF \bar{\Lambda}$, 
where $F$ and $\bar{\Lambda}$ were introduced below (\ref{eq1}),
and all seven matrix elements $\langle {\cal O}^{(5)}_k  \rangle$ for the dimension-5 
operators
can be expressed by $F$,
$\bar{\Lambda}$ and two additional HQET parameters $\lambda_E$ and $\lambda_H$,
which are associated with the
chromoelectric and chromomagnetic fields inside the $B$ meson, respectively,  as
$\langle \bar{q}g{\bf E\cdot\alpha}
\gamma_5 h_v \rangle
=F\lambda_E^2$ and
$\langle \bar{q}g{\bf H\cdot \sigma}
\gamma_5h_v \rangle
=i F
\lambda_H^2$ in the rest frame, $v=(1,{\bf 0})$.
We get the OPE form for the LCWF (\ref{eq1}) as 
\begin{eqnarray}
\tilde{\phi}_B(t,\mu)
&=&
iF(\mu)\left\{
1- \frac{\alpha_s C_F}{4\pi}
\left(2L^2+2L+\frac{5 \pi^2}{12}\right)
-it\frac{4\bar{\Lambda}}{3} 
\left[1- \frac{\alpha_s C_F}{4\pi}
\left(2L^2+4L-\frac{9}{4}+\frac{5\pi^2}{12} \right)
\right]
\right.
\nonumber\\
&&
\!\!\!\!\!\!\!\!\!\!\!\!\!\!\!\!\!\!\!\!\!\!\!\!\!\!\!\!\!\!\!\!\!\!\!
-t^2 \bar{\Lambda}^2\left[
1- \frac{\alpha_sC_F}{4\pi}
\left(2L^2+\frac{16}{3}L-\frac{35}{9}
+\frac{5\pi^2}{12} \right)
\right]
-\frac{t^2\lambda_E^2(\mu)}{3}
\left[1- \frac{\alpha_sC_F}{4\pi}
\left(2L^2+2L-\frac{2}{3}
+\frac{5\pi^2}{12} \right)
\right.
\nonumber\\
&&\!\!\!\!\!\!\!\!\!\!\!\!\!\!\!\!\!\!\!\!
\left.+ 
\frac{\alpha_sC_G}{4\pi}
\left(\frac{3}{4}L-\frac{1}{2}\right)
\right]
-\frac{t^2\lambda_H^2(\mu)}{6} 
\left.\left[1- \frac{\alpha_sC_F}{4\pi}
\left(2L^2+\frac{2}{3}
+\frac{5\pi^2}{12} \right)
-\frac{\alpha_s C_G}{8\pi}
\left(L-1\right)
\right]
\right\} ,
\label{eq4}
\end{eqnarray}
with the $\overline{\rm MS}$ scale $\mu$ and $\alpha_s \equiv \alpha_s(\mu)$,
which takes into account the Wilson coefficients 
to 
$O(\alpha_s)$ and a complete set of
the local operators of dimension less than 6.
Fourier transforming to the momentum 
%
representation
and taking the first two regularized-moments,
$\int_0^{\Lambda_{UV}} d\omega  \omega ^j \phi _B (\omega, \mu)$ ($j=0,1$),
the contributions from the first line in (\ref{eq4}), associated with matrix element of
the dimension-3 and -4 operators, coincide completely with
the result obtained in \cite{Lee:2005gza}.
The second and third lines in (\ref{eq4}) are
generated from the dimension-5 operators.
Our OPE result (\ref{eq4}) ``merges'' the IR and UV structures 
peculiar to the $B$-meson LCWF~\cite{KKQT,Lange:2003ff}, and
embodies novel behaviors that are completely different from those
of the pion LCWF:
$\mu$ and $t$ are strongly correlated due to the
logarithmic contributions from cusp singularity in the radiative corrections,
so that the WF is not Taylor expandable about $t=0$.
The WF receives the contributions from many higher dimensional operators,
in particular, from those associated with the long-distance gluon fields inside
$B$-meson.

Our result (\ref{eq4}) allows us to parameterize all nonperturbative contributions 
by a few HQET parameters.
As a result, 
the $B$-meson LCWF (\ref{eq1}) obeys the two-step evolution:
it is governed by
the Sudakov-type and Brodsky-Lepage-type scale dependence 
from the high scale $\mu_i\simeq\sqrt{m_b \Lambda_{\rm QCD}}$,
associated with the QCD factorization formula, to the scale $\mu \lesssim 1/t$,
while that for the lower scale is governed by the anomalous dimensions 
of the local operators associated with the HQET parameters.
%
This result is useful for clarifying 
model-independent properties 
of the $B$-meson LCWF,
combined with nonperturbative estimates of the relevant HQET parameters.

\bigskip
K.T. thanks V.~M. Braun 
for valuable
discussions. 
This work is supported by the Grant-in-Aid 
for Scientific Research No.~B-19340063.

\end{document}